\documentclass[final,5p,times,twocolumn]{elsarticle}

\usepackage{amsthm}
\usepackage{dblfloatfix}
\usepackage{euscript}   % for the calligraphic G
\usepackage{graphicx}% Include figure files
\usepackage{dcolumn}% Align table columns on the decimal point
\usepackage{bm}% bold math
\usepackage{amsmath}% bold math
\usepackage{amssymb}

\usepackage{textcomp}%Provides extra symbols, e.g. arrows like \textrightarrow, various currencies (\texteuro,...), things like \textcelsius and many others.
\usepackage{tabularx} % Ruled Tabular
\usepackage{times}
\usepackage{titlesec}
\usepackage{wrapfig}
\usepackage{caption}
\usepackage{subcaption}
\usepackage{adjustbox}
\usepackage{listings}
\usepackage{silence}
\usepackage{wasysym}
\usepackage{titlesec}
\usepackage{graphicx}
\usepackage{colortbl}
\usepackage[table,xcdraw]{xcolor}
\usepackage{booktabs}
\usepackage{soul}

\usepackage{graphicx}% Include figure files
\usepackage[utf8]{inputenc}
\usepackage[T1]{fontenc}
\usepackage{etoolbox}
\usepackage{csquotes}
\usepackage{xcolor}

\usepackage{hyperref}% add hypertext capabilities
\hypersetup{
    colorlinks=true,
    linkcolor=blue,
    filecolor=magenta,      
    urlcolor=cyan}

\usepackage{siunitx}
\usepackage{diagbox}

\begin{document}
\title{Defect Topology Governs the Optical Response of Doped Amorphous Silicon}

\author[OU,LANL]{R. Hussein}
\ead{rh353321@ohio.edu}
\affiliation[OU]{organization={Department of Physics and Astronomy},
            addressline={Ohio University}, 
            city={Athens},
            postcode={45701}, 
            state={OH},
            country={USA}}

\author[LANL]{C. Ugwumadu}
\ead{cugwumadu@lanl.gov}
\affiliation[LANL]{Quantum & Condensed Matter (T-4) Group, Los Alamos National Laboratory, Los Alamos, NM, USA}

\author[OX]{L. A. M. Rosset}
\ead{louise.rosset@chem.ox.ac.uk}
\affiliation[OX]{Inorganic Chemistry Laboratory, Department of Chemistry, University of Oxford, Oxford, UK}

\author[LANL]{R. M. Tutchton}
\ead{rtutchton@lanl.gov}

\begin{abstract}
 The influence of defect topology and chemical doping on the optical response of amorphous silicon ($a$-Si) remains unresolved. In this study, hybrid-functional density functional theory and linear-response calculations are employed to examine four structurally distinct continuous random networks with substitutional boron and phosphorus doping. The calculated optical properties reproduce key experimental trends and demonstrate that dopant sensitivity is determined not only by defect concentration, but also by the character and spatial organization of defects. A configuration with fully extended 4-fold coordination shows the strongest dielectric perturbation to doping, while dopant effects in networks with pre-existing defect states are more varied. For most networks, the dielectric response is more strongly perturbed by boron than phosphorus. Notably, a network having both 3- and 5-fold defects with a total defect fraction of $7.4\%$ reverses the typical boron-dominant response of other cases, becoming more sensitive to phosphorus. Tauc and optical sum-rule analyses further indicate that network topology defines the broader optical landscape, whereas doping primarily affects states near the gap and redistributes spectral weight. These findings establish defect topology as a microscopic control parameter that links amorphous structure, dopant activity, and optical response in amorphous silicon.
\end{abstract}

            
\date{}

\maketitle

%\section{Introduction}

Amorphous silicon ($a$-Si) is one of the most widely studied disordered semiconductors because of its technological importance and rich structure--property relationships. Unlike crystalline silicon, $a$-Si lacks long-range periodicity while largely preserving local tetrahedral coordination. Instead, its structure is that of the continuous random network (CRN), which contains bond-angle distortions, coordination irregularities, and topological disorder, all of which strongly influence the electronic density of states, the dielectric response, carrier transport and optical transitions \cite{Drabold_1990,MottDavisBook1979}. In particular, the absence of translational symmetry gives rise to localized band-tail states near the mobility edges \cite{Fedders-98-12,Drabold2028}, leading to optical-edge broadening and optical behavior that differs substantially from crystalline Si \cite{Dong,Atomistic_origin}.

Despite decades of work, generating realistic $a$-Si models that reproduce both structural and optoelectronic properties remain challenging. Different models may exhibit similar radial distribution functions, coordination statistics, and short-range order, yet still differ in topology, wavefunction localization, and electronic overlap \cite{Djordjevi,Barkema,Deringer}. These subtle differences can strongly affect the optical gap, dielectric screening, absorption edge, and sub-gap transitions. Defects such as dangling bonds, floating bonds, strained tetrahedral environments, and weakly bonded configurations are especially important because they modify band-tail and near-gap states, which in turn control optical absorption and carrier recombination \cite{Drabold_1990,Fedders,harke2010amorphous}.

The optical response of $a$-Si is also sensitive to sample preparation history and resulting structural changes, including thermal relaxation, hydrogen incorporation, density variations, and pressure \cite{GKM_Thutupalli_1977,Ravindra,Adachi,Duran}. Spectroscopic ellipsometry has shown this; $a-$Si dielectric spectra and optical constants depend on preparation method and structural state, with self-implanted and LP-CVD-grown samples differing in residual microcrystalline content \cite{Adachi}. The exponential absorption edge in amorphous semiconductors is commonly associated with disorder-induced band-tail states and described using the Urbach formalism, while higher-energy absorption is often analyzed using the Tauc model for transitions between extended states \cite{Ravindra}. These observations highlight the need to directly compare structure and optical properties across independently generated $a$-Si networks. 

\begin{figure*}[!t]
    \centering
    \includegraphics[width=\linewidth]{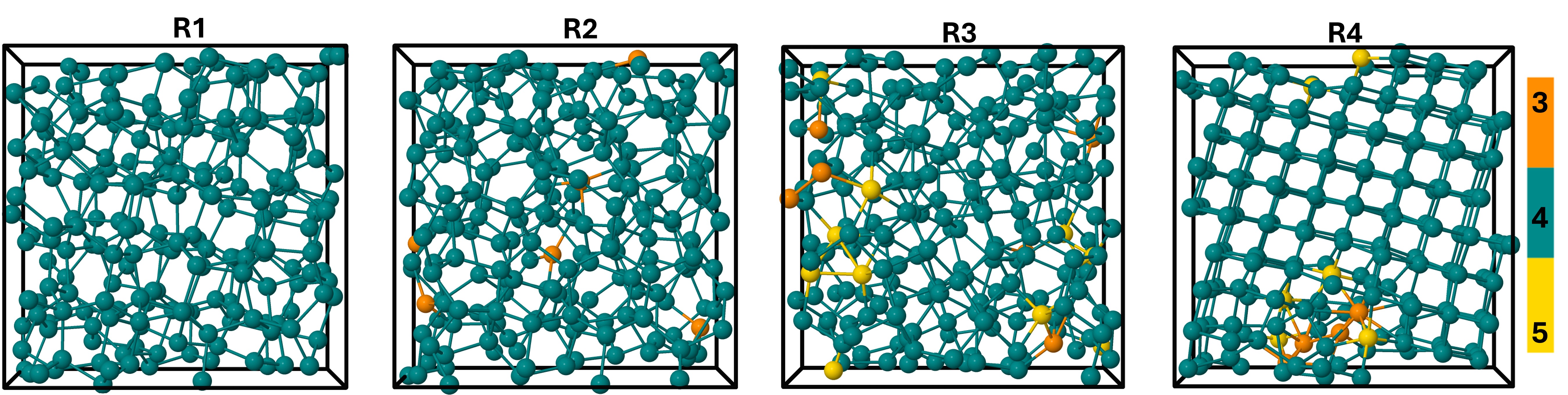}
    \caption{Structure of the $a-$Si models: R$1$, R$2$, R$3$ and R$4$. Side bar colored according to their local coordination number with Si-Si cutoff of $2.6$ \AA; three-fold (orange), four-fold (dark-cyan), five-fold (gold). R$4$ is a strained diamond-cubic sample with a cluster of defective atoms, unlike the dispersed defects in the amorphous R$2$ and R$3$ structures.}
    \label{fig:Model_doped}
\end{figure*}

Recent machine-learning molecular-dynamics simulations by Rosset \textit{et al.} \cite{Rosset2025-paracrystal-models} produced a dataset of high-quality disordered Si models generated through melt-quench protocols at varying quench rates and densities. Within this dataset, four structures were chosen, of which one is a defect-free $a$-Si model, \textit{`Rosset-10'} and hereafter R1, which exhibits a clean electronic gap and structural characteristics consistent with experiments \cite{Rosset2026}. Three additional models with different defect forms were also selected from the dataset, hereafter R2, R3 and R4.
In this work, we use first-principles linear-response calculations to study the optical properties of these four models, and how these properties respond to substitutional doping of boron and phosphorus. We compute the dielectric function, refractive index, extinction coefficient, absorption coefficient, reflectance, transmittance, and absorbance over a broad photon-energy range. We further obtain the Tauc gap and Urbach energy, thus overall clarifying how structural topology, native network disorder, and dopant-induced electronic modifications influence the optical response of $a$-Si at the atomistic level. This study provides the first investigation of the optical properties of the R1 model, a structurally distinct member of the defect-free amorphous silicon family \cite{Rosset2026}.

%\section{Computational Details}
First-principles calculations were performed within the density functional theory framework using the Vienna \textit{Ab initio} Simulation Package (VASP) \cite{VASP_0}. Interactions between valence electrons and ionic cores were described using the projector augmented-wave method \cite{PAW}, and exchange and correlation were treated using the Heyd--Scuseria--Ernzerhof hybrid functional (HSE06) \cite{HSE06}. The use of HSE06 is particularly relevant to the optical response of amorphous silicon because it improves the description of the band gap and the electronic states involved in optical transitions compared with commonly used DFT approximations such as the Perdew–Burke–Ernzerhof (PBE) functional \cite{PBE}. This is especially important near the absorption edge, where the response is sensitive to the energies and character of band-tail states. A plane-wave energy cutoff of \(520~\mathrm{eV}\) was used throughout, with Brillouin-zone sampling restricted to the \(\Gamma\) point. The electronic self-consistency convergence threshold was set to \(10^{-6}~\mathrm{eV}\), and structural relaxations were performed until the residual forces were below \(0.01~\mathrm{eV/\AA}\).

%\subsection{Structural Models}
The four structures exhibit distinct local coordination environments and spatial distribution of defects, as shown in Fig.~\ref{fig:Model_doped}. While R1 is defect-free, the models R2, R3, and R4 possess distinct native coordinative and topological differences.  The defective networks differ not only in the concentration but in their spatial orientation. R3 has the highest total defect fraction of 7.4\%, with R4 and R2 having 4.6 and 2.8\%, respectively. The coordination defects are spatially dispersed for R2 and R3, whereas the 3- and 5-fold defects in R4  are clustered. Substitutional doping was introduced into each network by replacing a single Si atom with either B or P. The doped structures were energetically relaxed prior to electronic structure and optical calculations. 

%\subsection{Optical response}
The optical response is described by the complex frequency-dependent dielectric function,
\begin{equation}
\varepsilon(\omega)=\varepsilon_1(\omega)+
i\varepsilon_2(\omega),
\end{equation}
which characterizes the interaction of electromagnetic radiation with the electronic structure \cite{palik1998handbook}. We computed the dielectric function from first-principles electronic-structure calculations within the independent-particle approximation (IPA). The imaginary part, \(\varepsilon_2(\omega)\), was obtained from interband transitions between occupied valence states and unoccupied conduction states, as in \cite{equations_1}:
\begin{equation}
\varepsilon_2(\omega) = \frac{4\pi^2 e^2}{\Omega} \lim_{q \to 0}
\frac{1}{q^2}\sum_{c,v,{k}} 
 \left\langle u_{c\bf{k}+e_{\alpha}{q}}| u_{v{k}} \right\rangle    \left\langle u_{c\bf{k}+e_\beta{q}}| u_{v{k}} \right\rangle 2 w_{{k}}\; \delta_{cv}.
\end{equation}
Here $e$ is the elementary charge, $\Omega$ the unit-cell volume, $u_{v,{k}}$ the cell-periodic block factor, $q$ is the wavevector magnitude,  \(w_{\mathbf k}\) is the \(\mathbf k\)-point weight. The delta function enforces energy conservation as 
\begin{equation}
    \delta_{cv} = \delta \left(E_c({k}) - E_v({k}) - \hbar \omega \right),
\end{equation}
where \(E_v\) and \(E_c\) are the valence- and conduction-band energies,

The real part, \(\varepsilon_1(\omega)\), was obtained from the Kramers$-$Kronig relation,
\begin{equation}
\varepsilon_1(\omega)=1+\frac{2}{\pi}\mathcal{P}\int_0^\infty
\frac{\varepsilon_2(\omega')\omega'}{\omega'^2-\omega^2}d\omega',
\end{equation}
where \(\mathcal{P}\) denotes the Cauchy principal value \cite{Linear_Optics}. The microscopic local-field effect associated with changes in the cell periodic part of the potential was neglected. Thus, the macroscopic optical response of the networks was obtained by averaging the 3 diagonal components of the dielectric tensor:
\begin{equation}
\varepsilon_{\mathrm{iso}}(\omega)
=
\frac{
\varepsilon_{xx}(\omega)
+
\varepsilon_{yy}(\omega)
+
\varepsilon_{zz}(\omega)
}{3}.
\end{equation}

The refractive index, $n(\omega)$ and extinction coefficient, $k(\omega)$, are obtained from the complex dielectric function as
\begin{equation}
n(\omega)=\left(\frac{\varepsilon_r+\varepsilon_1(\omega)}{2}\right)^\frac{1}{2}, \quad \text{and} \quad k(\omega)=\left(\frac{\varepsilon_r-\varepsilon_1(\omega)}{2}\right)^\frac{1}{2},
\end{equation}
respectively, where \(\varepsilon_r = \left[\varepsilon_1^2(\omega)+\varepsilon_2^2(\omega)\right]^\frac{1}{2}\). 

The absorption coefficient, $\alpha(\omega)$, is related to $k(\omega)$ by the relation
\begin{equation}
\alpha(\omega)=\frac{4\pi k(\omega)}{\lambda},
\end{equation}
and the optical conductivity is defined as
\begin{equation}
\sigma(\omega)=\varepsilon_0\omega\varepsilon_2(\omega).
\end{equation}

%\section{Results and Discussion}
\begin{figure}[!t]
    \centering
    \includegraphics[width=.95\linewidth]{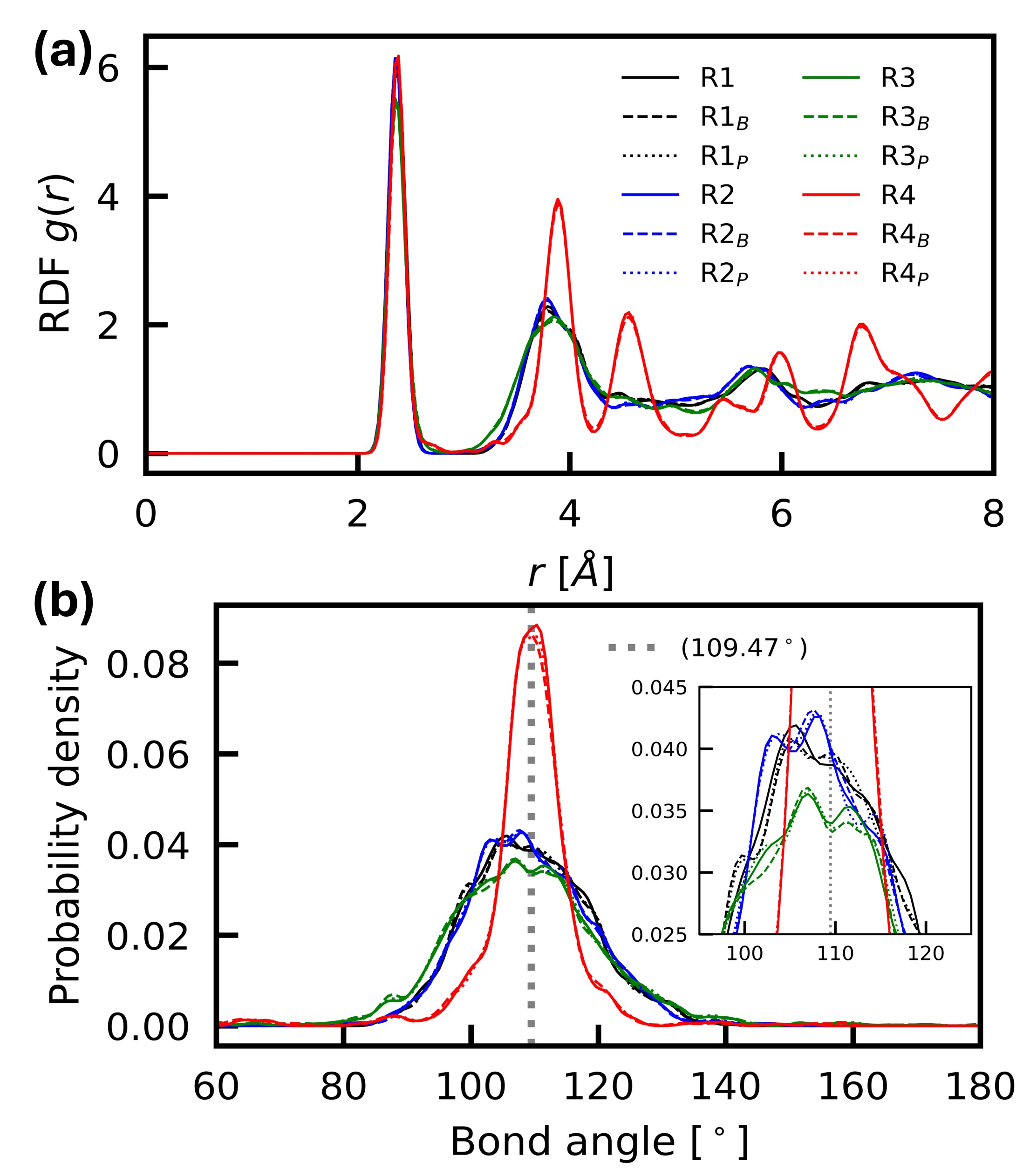}
    \caption{Radial distribution functions (RDF) $g(r)$, and (b) bond angle distributions (BAD) for the undoped structural models (R1$-$R4), and their B- and P-doped counterparts, in dashed and dotted lines respectively denoted with subscripts. Doped models closely overlap with their undoped variants. A dashed gray line in (b) indicates the ideal bond angle of 109.47$^{\circ}$.}
    \label{fig:Rdf_BAD}
\end{figure}

%Figure 3
\begin{figure*}[!t]
 \centering
	\includegraphics[width=\linewidth]{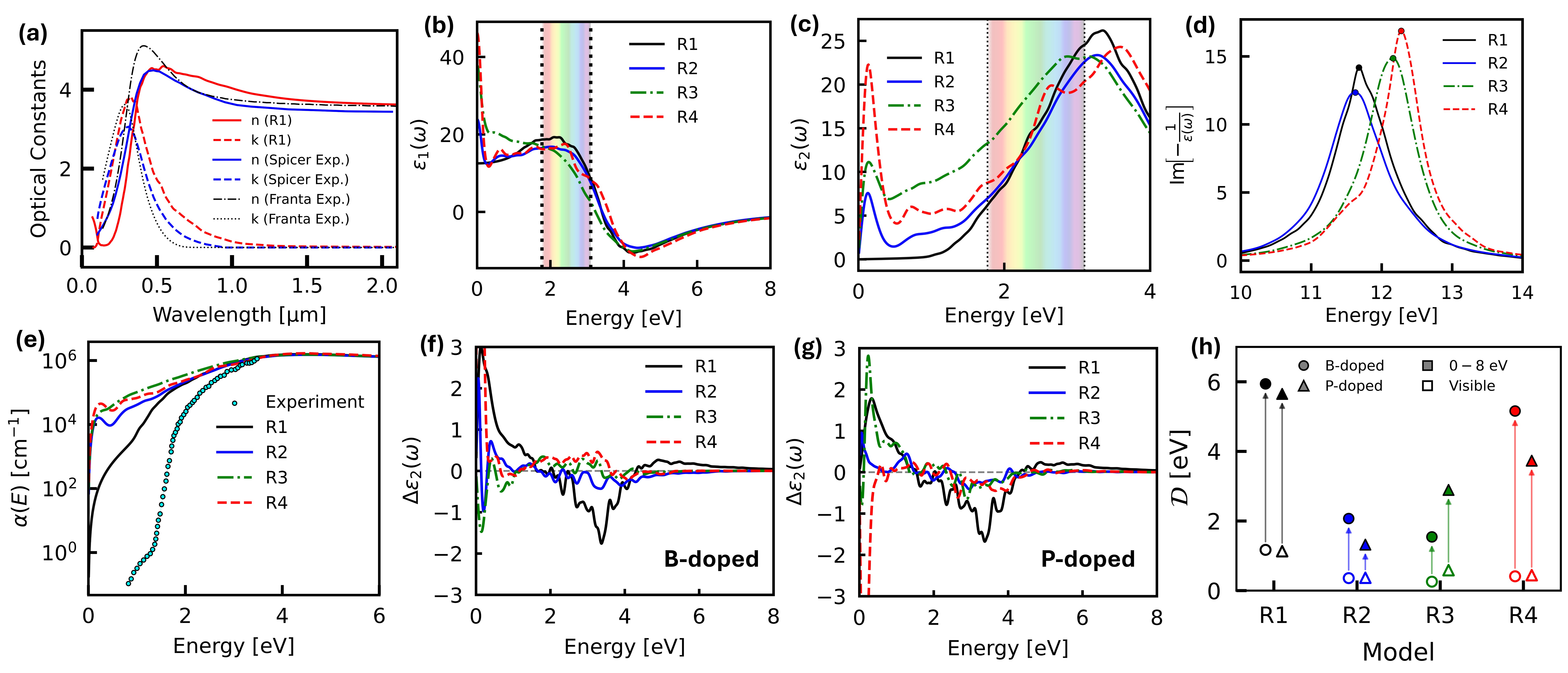}
	\caption{ (a) Refractive index $n(\omega)$ and extinction coefficients $k(\omega)$ of models as functions of wavelength, demonstrating good agreement between the defect-free model (R1) and experimental results from \cite{Pierce} and \cite{FRANTA201312}.  (b) Real $\varepsilon_{1}(\omega)$ and (c) imaginary $\varepsilon_{2}(\omega)$ components of the dielectric functions of the undoped structural models. The rainbow indicates the visible spectrum region. (d) Electron loss functions showing the peak for all undoped models. (e) Absorption coefficient of all four undoped models compared with experiment taken from \cite{Exp_absorption}. Change in imaginary component $\varepsilon_{2}(\omega)$ upon doping, for the (f) B-doped and (g) P- doped cases respectively. (h) Total change integrated over the entire energy range (0--8, filled) and visible (1.77--3.10, open) for B (circles and P (triangles).}
 \label{fig:Fig_n_k}
\end{figure*}

Figure \ref{fig:Rdf_BAD}(a) shows the first radial distribution function (RDF) with a nearest-neighbor peak at $2.35~\AA$ in all models, both doped and undoped. The second peak is broader for R1$-$R3 within $(r)=3.5-4.5$ \AA, and reduces afterwards as expected for CRNs. Comparatively, R4 displays a sharp second peak, followed by subsequent sharp peaks that persist to $r\approx 8$ \AA, for the accessible RDF range reflecting the structural order. As shown in Fig.~\ref{fig:Rdf_BAD}(b), the bond angle distribution (BAD) for R4 is very narrow at the ideal 109.47$^{\circ}$ while much broader for R1$-$R3. Bond-angle disorder has been correlated with the valence-band tail and, consequently, the Urbach edge in $a$-Si:H \cite{BUSTARRET198913}. Atomistic calculations also identify these distortions to the local tetrahedron as a contributing source of band-tail states in $a$-Si \cite{KANG2000169}; however, these states are more strongly associated with subnetworks of long and short bonds \cite{Fedders-98-12, Atomistic_origin, Inam-10, Drabold-11-01}. Further details of coordination numbers and defect fractions are available in \textcolor{blue}{Table S1} of the Supplemental Material (SM) \cite{SM}. Figure \ref{fig:Rdf_BAD} shows that substituting B and P dopants causes no significant change in the RDFs and BAD, especially for R1, R2, and R4, while R3 shows a subtle, dopant-dependent change in local bond-angle order. Thus, any optical change upon doping can be attributed predominantly to dopant-induced electronic perturbations of the network.

Figure \ref{fig:Fig_n_k} shows the relationship between the refractive index and the dielectric function. Figure \ref{fig:Fig_n_k}(a) plots the dependence of $n(\lambda)$ and $k(\lambda)$ on the wavelength ($\lambda$) for the defect-free R1 model. It suggests a decrease in $k$ as $\lambda$ increases, marking the onset of the optical gap as $k\to 0$. The overall trends agree well with experimental results \cite{Pierce,FRANTA201312}.

Figure \ref{fig:Fig_n_k}(b) shows that for $a$-Si, $\varepsilon_1(\omega)$ becomes negative in the photon-energy range of approximately $3$--$4$~eV and then rises toward a nearly constant value. The zero crossing of $\varepsilon_1(\omega)$ indicates strong electronic screening and enhanced reflectivity in this energy range \cite{negative}. Figure \ref{fig:Fig_n_k}(c) shows an important low-energy feature with R1 having negligible absorption below $1$~$eV$, in contrast to the other three models. This behavior is consistent with the increase in $\varepsilon_2(\omega)$ below the interband edge in amorphous materials, due to the presence of defects and thus, intra-band absorption \cite{Jellison}.

The electron energy-loss function $Im[-1/\varepsilon(\omega)]$ in Fig.~\ref{fig:Fig_n_k}(d), shows the collective coherent oscillations of the electron density (plasmons). The peaks span the $11.6-12.3$ $eV$ range, with R4 having largest shift ($+$0.6) relative to R$1$ with the highest peak intensity $\approx 16.9$. The loss function and optical conductivity of the doped models are shown in Fig.~\textcolor{blue}{S1} of the SM \cite{SM}.

Dopants perturb the networks differently, as shown in Fig.~\ref{fig:Fig_n_k}(f-h), with R1 being the most perturbed by both dopants. Defining
\begin{equation}
  \mathcal{D}=\int_{E_1}^{E_2}|\Delta\varepsilon_2(E)| dE,
\end{equation}
we obtain $\mathcal{D}=$ 6.0 and 5.7 eV for boron and phosphorus, respectively. As shown in Fig.~\ref{fig:Fig_n_k}(f) and (g), the persistent dip in $\Delta \varepsilon_2(\omega)$ in the $1.8-4.2$ eV range in R1 is not observed in R2, R3, and R4. With no pre-existing gap states in R1, the dopant-induced state lies within the empty gap and contributes its spectral weight to $\Delta \varepsilon_2(\omega)$. In the defective models, pre-existing dangling and floating bonds \cite{Pantelides} already occupy part of the same spectral window, so the dopant contribution is partly absorbed rather than added to the existing defect response, thus reducing the perturbation.

R4 exhibits $\mathcal{D}\approx 5.2$ eV and $3.7$ eV for B and P doping respectively; however, compared to R1, its $\Delta \varepsilon_2(\omega)$ is concentrated almost near $E\to 0$ rather than spread across the spectrum. Defect number alone does not account for these observed differences. R3 has the highest defect fraction, but gives $\mathcal{D}\approx 1.6$ eV (B) and 2.9 eV (P) , with  R2 given 2.1 eV (B) and 1.3 eV (P);  both smaller than R4. R3 exhibits anomalous behavior with P exceeding B response as compared to the pattern in R1, R2, and R4. This indicates that dopant perturbation of the network is not a fixed behavior of B or P doping, but depends on the specific defect topology of the host's network. A full table of $\mathcal{D}$ values is available in \textcolor{blue}{Table S2} of the SM \cite{SM}.

\begin{figure*}[!t]
 \centering
	\includegraphics[width=.9\linewidth]{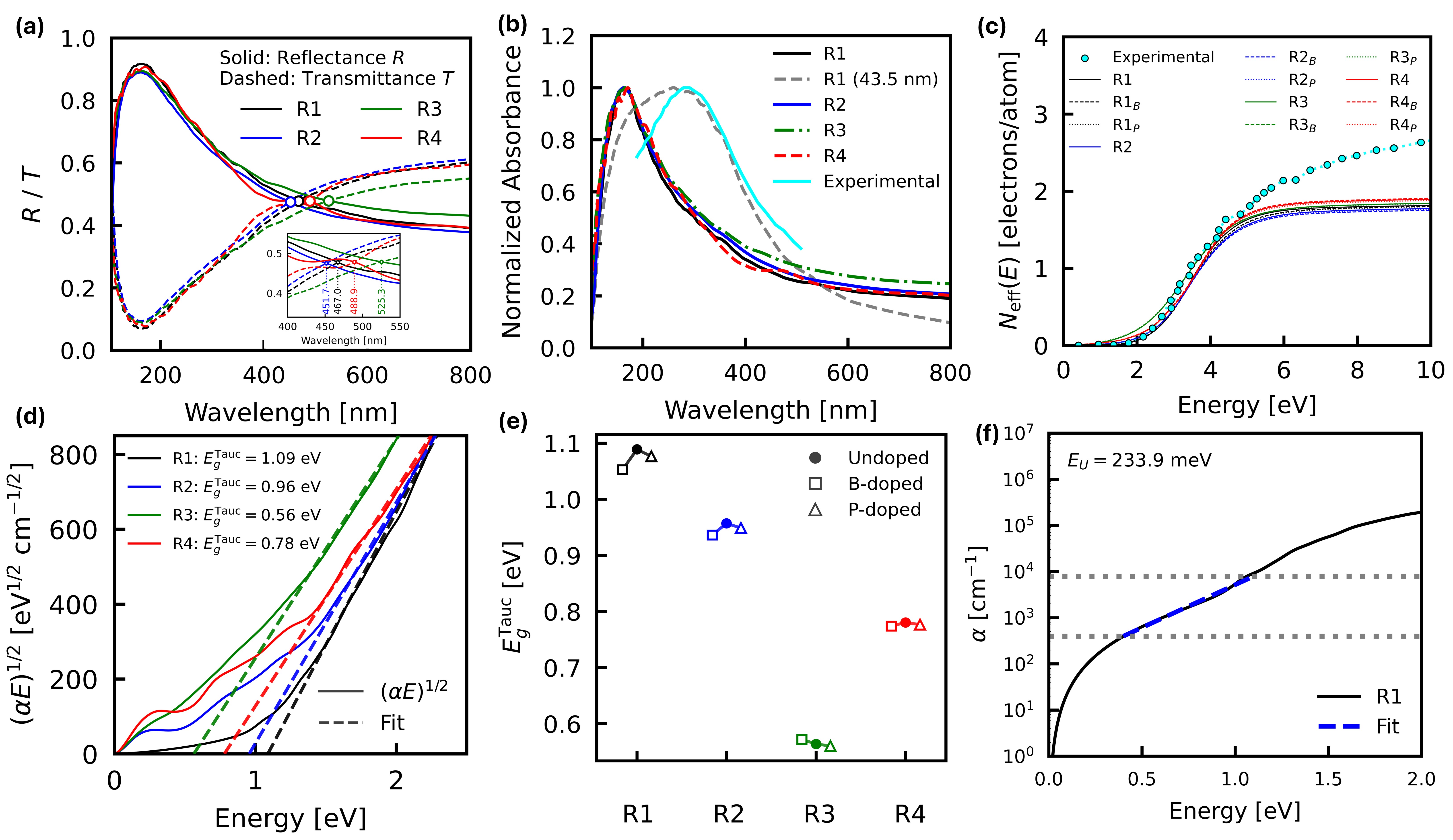}
	\caption{(a) Reflectance R (solid lines) and transmittance T (dashed lines) spectra of the four $a-$Si models as a function of the wavelength from the complex dielectric function. An inset shows the $R_{\lambda}=T_{\lambda}$ crossovers within the visible spectrum (between $452-525~nm$). (b) Normalized optical absorbance of the studied models compared with experimental data \cite{Deep_UV}. (c) Effective electron number of all pristine and doped configurations with an experimental plot from Ref. \cite{MottDavisBook1979} with energy. (d) Tauc plots, $(\alpha E)^{1/2}$ vs photon energy, with linear fits used to determine the optical gap for each model. (e) Summary of Tauc plots of undoped and doped R1$-$R4. (f) Urbach plot as a function of photon energy for undoped R1. The linear region below the optical absorption edge is fitted to extract the Urbach energy $E_U$. }
 \label{fig:loss_cond_alpha}
\end{figure*}

The reflectance $R(\omega)$ and transmittance $T(\omega)$  were evaluated using the Beer-Lambert law, joined with Fresnel's reflection loss equations as shown in Fig.~\ref{fig:loss_cond_alpha}(a). Both equations combine to account for the effects of reflection losses and absorption of the materials, and  are expressed as \cite{brunin2019transparent}
\begin{align}
    R(\omega)&=\frac{(n-1)^2+k^2)}{(n+1)^2+k^2},\\
    T(\omega)&=(1-R(\omega))e^{{-\alpha (\omega)d}}.
\end{align}
The effective optical path $d_{eff}=l$ is the simulation cell edge for each model's  cell cubic,$(l_{x}=l_{y}=l_{z})$. The edge lengths are $16.548$, $16.525$, $16.237$ and $16.324$ $\AA$ for R1$-$R4, and for both their B- and P- doped models. $R(\omega)$ and $T(\omega)$ spectra of all four undoped models are similar in the deep UV range, with a peak $R(\omega)\approx 0.9$ near $150$ nm. 
The wavelength where $R=T$ differs between models, taking values of $\approx 452, 467$, $523$ and $489$ nm for models R1 through R4 respectively. The higher crossover value for R3 is associated with its defect topology. A noticeable feature is the non-monotonic behavior of R4 near the crossover, which we attribute to the predominantly diamond-cubic lattice with the localized defective domain. Thus, a crystalline-like and defect-localization contribution rather than a single amorphous response from R1-R3. Figure \textcolor{blue}{S2} \cite{SM} shows plots for $A$, $R$ and $T$ for all the doped models.

%FIGURE 4

We obtained the spectroscopic absorbance for the four undoped models, calculated as 
\begin{equation}
 \mathrm{Abs}(\omega)=-\mathrm{log_{10}}(T)
\end{equation}
and shown in Fig.~\ref{fig:loss_cond_alpha}(b) with comparison to experimental UV-visible absorbance spectra \cite{Deep_UV}. The calculated spectra show a narrower peak than that from experiment; thus, to assess the effect of film thickness, the R1 absorbance was recalculated using the experimental thin-film thickness of $43.5$ nm (gray-dashed) which improves the peak agreement with experiment data. The remaining difference in peak position reflects effects beyond IPA and structural heterogeneity present in the experimental film. 

We further analyzed the effective number of electrons participating in the optical transitions for each model. $N_{eff}$ is obtained using the optical sum rule \cite{MottDavisBook1979}:
 \begin{equation}
     N_{eff}(E)=\frac{2m_{e}\varepsilon_0 V}{\pi e^2 N}\int_{0}^{\omega(E)} \omega' \,\varepsilon_2(\omega') d\omega',
 \end{equation} 
where $\varepsilon$ is the permittivity of vacuum, $m_e$ is the mass of the electron, $e$ electron charge and $N$ number of atoms. This yields the spectral weight of the optical transitions, interpreted as the effective number of electrons per atom that eventually contribute to the absorption up to a given photon energy. As shown in Fig.~\ref{fig:loss_cond_alpha}(c), $N_{eff}$ goes to zero at low photon energies, indicating the absence of any allowed transitions within the gap. At 10 eV, $N_{eff}$ values for the undoped models vary by 7\% (low-high), and doping shifts $N_{eff}$ by at most 1.4\%. P-doping decreases $N_{eff}$ in each model $(-0.2$ to $0.9\%)$, while B-doping causes a decrease for R1 and R2 $(-0.4$ to $ -1.35\%)$  but a small increase for R3 and R4 $(0.05,0.7\%)$. 
The reduction in $N_{eff}$ may be associated with the localized character of the electrons as described by Bauer in Ref.~\cite{MottDavisBook1979}.

We further our analysis by extrapolating the linear fit of ($\alpha E$)$^{1/ \gamma}$ \cite{PRASADARAO20094579} with the photon energy, obtaining the optical gap $E_g^{{Tauc}}$ for each model using $\gamma=2$ exponent for the indirect transition allowed gap in silicon \cite{TAUC196837}. The obtained gaps track differently for different topological behavior with defects. R1 has the highest gap $ \approx 1.1~{eV}$; R2 and R4 have 0.96 and 0.78 eV respectively, and R3, which contains higher defects, has the smallest gap of 0.56 eV, where it cuts the energy axis as shown in Fig.~\ref{fig:loss_cond_alpha}(d). The R1 result is consistent with that obtained by ~\cite{similar}. The linear fit quality is similar across all cases with $R^2\sim0.993$. Also, the doping cases showed an ordering of R$1>$R$2>$R$4>$R$3$ similar to the undoped cases, as shown in the plot of the optical gaps [Fig.~\ref{fig:loss_cond_alpha}(e)]. However, peculiar behavior of B-doped lowering the gap relative to P-doped cases, except R3, was observed. 

By investigating the sub-gap absorption behavior of R1 and the other models, we analyzed the Urbach tail from the absorption spectrum 
\text{$
    \alpha(E)=\alpha_{0} \exp\left(\frac{E}{E_U}\right) 
$} \cite{Urbach_Energy} and then using the linear relation,
\begin{equation}
    \ln\alpha(E)=\ln(\alpha_{0})+ \frac{E}{E_U} .
\end{equation} 

Figure \ref{fig:loss_cond_alpha}(f) shows $\alpha(\hbar\omega)$ as a function of photon energy, where $\alpha_0$ is a constant depending on the material and $E_U$ is the Urbach energy. For the Urbach-tail analysis, we used the absorption spectrum between $(500\leq\alpha(\hbar\omega)\leq 5000)$ $cm^{-1}$ and energy $(0.4\leq E \leq 1.1)$ $[eV]$. The same energy range was used for all models (see Fig.~\textcolor{blue}{S3} \cite{SM}) in the SM for completeness. This regime was employed in earlier $a-$Si works since Cody's foundational work on $a-$Si:H \cite{PERSANS1984203,SWEENOR1999281}. The linear least-squares fit of $\ln(\alpha)$ gives $233.9$ meV, with COD $R^2=0.993$, showing a good fit. From slope $1/E_U$, which characterizes the exponential band tail \cite{Urbach_Tail} is consistent with values reported for hydrogen-free $a-$Si; 229 meV for MBE-growth material \cite{OLEARY200157} and $130-356$ meV across sputtered films \cite{coatings12101549} and within the $170-240$ meV reported for ion bombardment $a-$Si \cite{170_240} which depends on the degree of structural disorder and the conditions used in the preparation of that sample. Structural motifs such as topological filaments and bond length distortions are reported, with $E_U=107$ meV ~\cite{Atomistic_origin}.

%\section{Conclusion}
In summary, this work investigates the optical response of amorphous silicon to substitutional doping, as a function of defect type and topology. By comparing 4 structurally distinct 216-atom networks (defect-free, dispersed 3-fold, 3 plus 5-fold, and clustered defects forming a domain), the defect-free model R1, exhibits the strongest dopant-induced dielectric perturbation with no gap states. In models with defects, that is, R2, R3 and R4, the dopant competes with the existing defect states and redistributes the spectral strength.  In contrast, R3, which has the highest defect fraction with dispersed  3- and 5-fold coordination defects,  exhibits a stronger dielectric response to phosphorus than to boron, reversing the B-doped dominant trend observed in R1, R2, and R4. Tauc analysis and effective electron numbers show that it is the distinct underlying structural topology, not the dopant species, that predominantly affects the bulk optical properties, even as the near-gap response depends strongly on the dopant. Doping was shown to alter the accumulated electrons by a small amount. These results establish defect topology as a microscopic control variable, linking structural disorder, impurity states, and optical response in disordered phases of silicon.

\section*{Funding}
This work was funded by the Laboratory Directed Research and Development program at Los Alamos National Laboratory through a project No.: 20240397ER, 20261284IR, and a Director’s Postdoctoral Fellowship (20240877PRD4). Los Alamos National Laboratory is operated by Triad National Security, LLC, for the U.S. DOE National Nuclear Security Administration (89233218CNA000001).

\bibliographystyle{elsarticle-num}
\bibliography{main_ref}

\begin{thebibliography}{10}
\expandafter\ifx\csname url\endcsname\relax
  \def\url#1{\texttt{#1}}\fi
\expandafter\ifx\csname urlprefix\endcsname\relax\def\urlprefix{URL }\fi
\expandafter\ifx\csname href\endcsname\relax
  \def\href#1#2{#2} \def\path#1{#1}\fi

\bibitem{Drabold_1990}
D.~A. Drabold, P.~A. Fedders, O.~F. Sankey, J.~D. Dow, Molecular-dynamics simulations of amorphous si, Phys. Rev. B 42 (1990) 5135--5141.
\newblock \href {https://doi.org/10.1103/PhysRevB.42.5135} {\path{doi:10.1103/PhysRevB.42.5135}}.

\bibitem{MottDavisBook1979}
N.~F. Mott, E.~A. Davis, Theory of electrons in a non-crystalline medium, in: Electronic Processes in Non-crystalline Materials, 2nd Edition, Clarendon/Oxford University Press, Oxford, New York, 1979, Ch.~2, pp. 11--15.

\bibitem{Fedders-98-12}
P.~A. Fedders, D.~A. Drabold, S.~Nakhmanson, Theoretical study on the nature of band-tail states in amorphous {Si}, Phys.~Rev.~B 58~(23) (1998) 15624--15631.
\newblock \href {https://doi.org/10.1103/PhysRevB.58.15624} {\path{doi:10.1103/PhysRevB.58.15624}}.

\bibitem{Drabold2028}
D.~A. Drabold, What atomistic studies of amorphous silicon have taught us about disorder, unpublished.

\bibitem{Dong}
J.~Dong, D.~A. Drabold, Atomistic structure of band-tail states in amorphous silicon, Phys. Rev. Lett. 80 (1998) 1928--1931.
\newblock \href {https://doi.org/10.1103/PhysRevLett.80.1928} {\path{doi:10.1103/PhysRevLett.80.1928}}.

\bibitem{Atomistic_origin}
Y.~Pan, F.~Inam, M.~Zhang, D.~A. Drabold, Atomistic origin of urbach tails in amorphous silicon, Phys. Rev. Lett. 100 (2008) 206403.
\newblock \href {https://doi.org/10.1103/PhysRevLett.100.206403} {\path{doi:10.1103/PhysRevLett.100.206403}}.

\bibitem{Djordjevi}
B.~R. Djordjevi\'{c}, M.~F. Thorpe, F.~Wooten, Computer model of tetrahedral amorphous diamond, Phys. Rev. B 52 (1995) 5685--5689.
\newblock \href {https://doi.org/10.1103/PhysRevB.52.5685} {\path{doi:10.1103/PhysRevB.52.5685}}.

\bibitem{Barkema}
G.~T. Barkema, N.~Mousseau, High-quality continuous random networks, Phys. Rev. B 62 (2000) 4985--4990.
\newblock \href {https://doi.org/10.1103/PhysRevB.62.4985} {\path{doi:10.1103/PhysRevB.62.4985}}.

\bibitem{Deringer}
V.~L. Deringer, N.~Bernstein, A.~P. Bartok, M.~J. Cliffe, R.~N. Kerber, L.~E. Marbella, C.~P. Grey, S.~R. Elliott, G.~Csanyi, Realistic atomistic structure of amorphous silicon from machine-learning-driven molecular dynamics, The Journal of Physical Chemistry Letters 9~(11) (2018) 2879--2885.
\newblock \href {https://doi.org/10.1021/acs.jpclett.8b00902} {\path{doi:10.1021/acs.jpclett.8b00902}}.

\bibitem{Fedders}
P.~A. Fedders, D.~A. Drabold, P.~Ordej\'on, G.~Fabricius, D.~S\'anchez-Portal, E.~Artacho, J.~M. Soler, Application of local-spin-density approximation to $a\ensuremath{-}\mathrm{Si}$ and tetrahedral $a\ensuremath{-}\mathrm{C}$, Phys. Rev. B 60 (1999) 10594--10597.
\newblock \href {https://doi.org/10.1103/PhysRevB.60.10594} {\path{doi:10.1103/PhysRevB.60.10594}}.

\bibitem{harke2010amorphous}
A.~Harke, \href{https://d-nb.info/1001371852/34}{Amorphous silicon for the application in integrated optics}, Ph.D. thesis, Hamburg University of Technology (2010).
\newline\urlprefix\url{https://d-nb.info/1001371852/34}

\bibitem{GKM_Thutupalli_1977}
G.~K.~M. Thutupalli, S.~G. Tomlin, The optical properties of amorphous and crystalline silicon, Journal of Physics C: Solid State Physics 10~(3) (1977) 467.
\newblock \href {https://doi.org/10.1088/0022-3719/10/3/017} {\path{doi:10.1088/0022-3719/10/3/017}}.

\bibitem{Ravindra}
N.~Ravindra, J.~Narayan, Optical properties of amorphous silicon and silicon dioxide, Journal of Applied Physics 60 (1986) 1139--1146.
\newblock \href {https://doi.org/10.1063/1.337358} {\path{doi:10.1063/1.337358}}.

\bibitem{Adachi}
S.~Adachi, H.~Mori, Optical properties of fully amorphous silicon, Phys. Rev. B 62 (2000) 10158--10164.
\newblock \href {https://doi.org/10.1103/PhysRevB.62.10158} {\path{doi:10.1103/PhysRevB.62.10158}}.

\bibitem{Duran}
M.~Durandurdu, D.~A. Drabold, Ab initio simulation of first-order amorphous-to-amorphous phase transition of silicon, Phys. Rev. B 64 (2001) 014101.
\newblock \href {https://doi.org/10.1103/PhysRevB.64.014101} {\path{doi:10.1103/PhysRevB.64.014101}}.

\bibitem{Rosset2025-paracrystal-models}
L.~A.~M. Rosset, D.~A. Drabold, V.~L. Deringer, Signatures of paracrystallinity in amorphous silicon from machine-learning-driven molecular dynamics, Nature Communications 16~(1) (Mar. 2025).
\newblock \href {https://doi.org/10.1038/s41467-025-57406-4} {\path{doi:10.1038/s41467-025-57406-4}}.

\bibitem{Rosset2026}
L.~A.~M. Rosset, C.~Ugwumadu, S.~R. Elliott, D.~A. Drabold, V.~L. Deringer, Defect-free model of amorphous silicon with pristine electronic structure, Phys. Rev. B (Aug 2026).
\newblock \href {https://doi.org/10.1103/cvf3-cd4p} {\path{doi:10.1103/cvf3-cd4p}}.

\bibitem{VASP_0}
G.~Kresse, J.~Furthm\"uller, Efficient iterative schemes for ab initio total-energy calculations using a plane-wave basis set, Phys. Rev. B 54 (1996) 11169--11186.
\newblock \href {https://doi.org/10.1103/PhysRevB.54.11169} {\path{doi:10.1103/PhysRevB.54.11169}}.

\bibitem{PAW}
P.~E. Bl\"ochl, Projector augmented-wave method, Phys. Rev. B 50 (1994) 17953--17979.
\newblock \href {https://doi.org/10.1103/PhysRevB.50.17953} {\path{doi:10.1103/PhysRevB.50.17953}}.

\bibitem{HSE06}
J.~Heyd, G.~E. Scuseria, M.~Ernzerhof, Hybrid functionals based on a screened coulomb potential, The Journal of Chemical Physics 118~(18) (2003) 8207--8215.
\newblock \href {https://doi.org/10.1063/1.1564060} {\path{doi:10.1063/1.1564060}}.

\bibitem{PBE}
J.~P. Perdew, K.~Burke, M.~Ernzerhof, Generalized gradient approximation made simple, Phys. Rev. Lett. 77 (1996) 3865--3868.
\newblock \href {https://doi.org/10.1103/PhysRevLett.77.3865} {\path{doi:10.1103/PhysRevLett.77.3865}}.

\bibitem{palik1998handbook}
E.~Palik, Handbook of Optical Constants of Solids, no. v. 3 in Academic Press handbook series, Elsevier Science, 1998.

\bibitem{equations_1}
M.~Gajdoš, K.~Hummer, G.~Kresse, J.~Furthmüller, F.~Bechstedt, Linear optical properties in the projector-augmented wave methodology, Physical Review B 20 (01 2006).
\newblock \href {https://doi.org/10.1103/PhysRevB.73.045112} {\path{doi:10.1103/PhysRevB.73.045112}}.

\bibitem{Linear_Optics}
M.~Gajdo\ifmmode~\check{s}\else \v{s}\fi{}, K.~Hummer, G.~Kresse, J.~Furthm\"uller, F.~Bechstedt, Linear optical properties in the projector-augmented wave methodology, Phys. Rev. B 73 (2006) 045112.
\newblock \href {https://doi.org/10.1103/PhysRevB.73.045112} {\path{doi:10.1103/PhysRevB.73.045112}}.

\bibitem{Pierce}
D.~T. Pierce, W.~E. Spicer, Electronic structure of amorphous si from photoemission and optical studies, Phys. Rev. B 5 (1972) 3017--3029.
\newblock \href {https://doi.org/10.1103/PhysRevB.5.3017} {\path{doi:10.1103/PhysRevB.5.3017}}.

\bibitem{FRANTA201312}
D.~Franta, D.~Nečas, L.~Zajíčková, I.~Ohlídal, J.~Stuchlík, Advanced modeling for optical characterization of amorphous hydrogenated silicon films, Thin Solid Films 541 (2013) 12--16, current Trends in Optical and X-Ray Metrology of Advanced Materials for Nanoscale Devices III.
\newblock \href {https://doi.org/https://doi.org/10.1016/j.tsf.2013.04.129} {\path{doi:https://doi.org/10.1016/j.tsf.2013.04.129}}.

\bibitem{Exp_absorption}
A.~Feltrin, R.~Bartlome, C.~Battaglia, M.~Boccard, G.~Bugnon, P.~Bühlmann, O.~Cubero, M.~Despeisse, D.~Domine, F.-J. Haug, F.~Meillaud, X.~Niquille, G.~Parascandolo, T.~Söderström, B.~Strahm, V.~Terrazzoni, N.~Wyrsch, C.~Ballif, An introduction to the technology of thin film silicon photovoltaics, Informacije MIDEM 39 (12 2009).

\bibitem{BUSTARRET198913}
E.~Bustarret, Urbach edges and angular distortions in a-si:h and related alloys, Journal of Non-Crystalline Solids 114 (1989) 13--15, proceedings of the Thirteenth International Conference on Amorphous and Liquid Semiconductors.
\newblock \href {https://doi.org/https://doi.org/10.1016/0022-3093(89)90054-9} {\path{doi:https://doi.org/10.1016/0022-3093(89)90054-9}}.

\bibitem{KANG2000169}
H.~Kang, Correlation between bond distortion and the band-tail electronic density of states in amorphous silicon: a tight-binding recursion study, Journal of Non-Crystalline Solids 261~(1) (2000) 169--180.
\newblock \href {https://doi.org/https://doi.org/10.1016/S0022-3093(99)00592-X} {\path{doi:https://doi.org/10.1016/S0022-3093(99)00592-X}}.

\bibitem{Inam-10}
F.~Inam, J.~P. Lewis, D.~A. Drabold, Hidden structure in amorphous solids, Phys.~Status Solidi A 207~(3) (2010) 599--604.
\newblock \href {https://doi.org/https://doi.org/10.1002/pssa.200982877} {\path{doi:https://doi.org/10.1002/pssa.200982877}}.

\bibitem{Drabold-11-01}
D.~A. Drabold, Y.~Li, B.~Cai, M.~Zhang, Urbach tails of amorphous silicon, Phys.~Rev.~B 83~(4) (2011) 045201.
\newblock \href {https://doi.org/10.1103/PhysRevB.83.045201} {\path{doi:10.1103/PhysRevB.83.045201}}.

\bibitem{SM}
See Supplemental Material at [URL inserted by publisher] for additional details on this work.

\bibitem{negative}
C.~Li, Z.~Wang, 9 - computational modelling and ab initio calculations in max phases – i, in: I.~Low (Ed.), Advances in Science and Technology of Mn+1AXn Phases, Woodhead Publishing, 2012, pp. 197--222.
\newblock \href {https://doi.org/https://doi.org/10.1533/9780857096012.197} {\path{doi:https://doi.org/10.1533/9780857096012.197}}.

\bibitem{Jellison}
J.~Jellison, G.~E., F.~A. Modine, Parameterization of the optical functions of amorphous materials in the interband region, Applied Physics Letters 69~(3) (1996) 371--373.
\newblock \href {https://doi.org/10.1063/1.118064} {\path{doi:10.1063/1.118064}}.

\bibitem{Pantelides}
S.~T. Pantelides, Defects in amorphous silicon: A new perspective, Phys. Rev. Lett. 57 (1986) 2979--2982.
\newblock \href {https://doi.org/10.1103/PhysRevLett.57.2979} {\path{doi:10.1103/PhysRevLett.57.2979}}.

\bibitem{brunin2019transparent}
G.~Brunin, F.~Ricci, V.-A. Ha, G.-M. Rignanese, G.~Hautier, Transparent conducting materials discovery using high-throughput computing, npj Computational Materials 5~(1) (2019) 63.
\newblock \href {https://doi.org/https://doi.org/10.1038/s41524-019-0200-5} {\path{doi:https://doi.org/10.1038/s41524-019-0200-5}}.

\bibitem{Deep_UV}
C.~H. Fields, W.~G. Oldham, R.~J. Bojko, {Use of amorphous silicon for deep-UV masks}, in: J.~D. Cuthbert (Ed.), Optical/Laser Microlithography, Vol. 1927, International Society for Optics and Photonics, SPIE, 1993, pp. 727 -- 735.
\newblock \href {https://doi.org/10.1117/12.150467} {\path{doi:10.1117/12.150467}}.

\bibitem{PRASADARAO20094579}
T.~{Prasada Rao}, M.~Santhoshkumar, Effect of thickness on structural, optical and electrical properties of nanostructured zno thin films by spray pyrolysis, Applied Surface Science 255~(8) (2009) 4579--4584.
\newblock \href {https://doi.org/https://doi.org/10.1016/j.apsusc.2008.11.079} {\path{doi:https://doi.org/10.1016/j.apsusc.2008.11.079}}.

\bibitem{TAUC196837}
J.~Tauc, Optical properties and electronic structure of amorphous ge and si, Materials Research Bulletin 3~(1) (1968) 37--46.
\newblock \href {https://doi.org/https://doi.org/10.1016/0025-5408(68)90023-8} {\path{doi:https://doi.org/10.1016/0025-5408(68)90023-8}}.

\bibitem{similar}
M.~Legesse, M.~Nolan, G.~Fagas, Revisiting the dependence of the optical and mobility gaps of hydrogenated amorphous silicon on hydrogen concentration, The Journal of Physical Chemistry C 117~(45) (2013) 23956--23963.
\newblock \href {https://doi.org/10.1021/jp408414f} {\path{doi:10.1021/jp408414f}}.

\bibitem{Urbach_Energy}
C.~Zhang, J.~Yuan, J.~K.~W. Ho, J.~Song, H.~Zhong, Y.~Xiao, W.~Liu, X.~Lu, Y.~Zou, S.~so, Correlating the molecular structure of a‐da'd‐a type non‐fullerene acceptors to its heat transfer and charge transport properties in organic solar cells, Advanced Functional Materials 31 (2021) 2101627.
\newblock \href {https://doi.org/10.1002/adfm.202101627} {\path{doi:10.1002/adfm.202101627}}.

\bibitem{PERSANS1984203}
P.~Persans, A.~Ruppert, S.~Chan, G.~Cody, Relationship between bond angle disorder and the optical edge of a-ge:h, Solid State Communications 51~(4) (1984) 203--207.
\newblock \href {https://doi.org/https://doi.org/10.1016/0038-1098(84)90996-7} {\path{doi:https://doi.org/10.1016/0038-1098(84)90996-7}}.

\bibitem{SWEENOR1999281}
D.~E. Sweenor, S.~K. O’Leary, B.~E. Foutz, On defining the optical gap of an amorphous semiconductor: an empirical calibration for the case of hydrogenated amorphous silicon, Solid State Communications 110~(5) (1999) 281--286.
\newblock \href {https://doi.org/https://doi.org/10.1016/S0038-1098(99)00049-6} {\path{doi:https://doi.org/10.1016/S0038-1098(99)00049-6}}.

\bibitem{Urbach_Tail}
D.~Drabold, Y.~Li, B.~Cai, M.~Zhang, Urbach tails of amorphous silicon, PHYSICAL REVIEW B 83 (01 2011).
\newblock \href {https://doi.org/10.1103/PhysRevB.83.045201} {\path{doi:10.1103/PhysRevB.83.045201}}.

\bibitem{OLEARY200157}
Optical dispersion relationships in amorphous silicon grown by molecular beam epitaxy, Journal of Non-Crystalline Solids 290~(1) (2001) 57--63.
\newblock \href {https://doi.org/https://doi.org/10.1016/S0022-3093(01)00728-1} {\path{doi:https://doi.org/10.1016/S0022-3093(01)00728-1}}.

\bibitem{coatings12101549}
M.~Ballester, M.~García, A.~P. Márquez, E.~Blanco, S.~M. Fernández, D.~Minkov, A.~K. Katsaggelos, O.~Cossairt, F.~Willomitzer, E.~Márquez, Application of the holomorphic tauc-lorentz-urbach function to extract the optical constants of amorphous semiconductor thin films, Coatings 12~(10) (2022).
\newblock \href {https://doi.org/10.3390/coatings12101549} {\path{doi:10.3390/coatings12101549}}.

\bibitem{170_240}
U.~Zammit, F.~Gasparrini, M.~Marinelli, R.~Pizzoferrato, A.~Agostini, F.~Mercuri, Ion dose effect in subgap absorption spectra of defects in ion implanted gaas and si, Journal of Applied Physics 70~(11) (1991) 7060--7064.
\newblock \href {https://doi.org/10.1063/1.349785} {\path{doi:10.1063/1.349785}}.

\end{thebibliography}
\end{document}

\title{SUPPLEMENTARY MATERIAL \\{Defect Topology Governs the Optical Response of Doped Amorphous Silicon}}

\author[OU,LANL]{R. Hussein}
\ead{rh353321@ohio.edu}
\affiliation[OU]{organization={Department of Physics and Astronomy},
            addressline={Ohio University}, 
            city={Athens},
            postcode={45701}, 
            state={OH},
            country={USA}}
\affiliation[LANL]{Quantum & Condensed Matter (T-4) Group, Los Alamos National Laboratory, Los Alamos, NM, USA}

\author[LANL]{C. Ugwumadu}
\ead{cugwumadu@lanl.gov}

\author[OX]{L. A. M. Rosset}
\ead{louise.rosset@chem.ox.ac.uk}
\affiliation[OX]{Inorganic Chemistry Laboratory, Department of Chemistry, University of Oxford, Oxford, UK}

\author[LANL]{R. M. Tutchton}
\ead{rtutchton@lanl.gov}

\begin{abstract}
 The influence of defect topology and chemical doping on the optical response of amorphous silicon ($a$-Si) remains unresolved. In this study, hybrid-functional density functional theory and linear-response calculations are employed to examine four structurally distinct continuous random networks with substitutional boron and phosphorus doping. The calculated optical properties reproduce key experimental trends and demonstrate that dopant sensitivity is determined not only by defect concentration, but also by the character and spatial organization of defects. A configuration with fully extended 4-fold coordination shows the strongest dielectric perturbation to doping, while dopant effects in networks with pre-existing defect states are more varied. For most networks, the dielectric response is more strongly perturbed by boron than phosphorus. Notably, a network having both 3- and 5-fold defects with a total defect fraction of $7.4\%$ reverses the typical boron-dominant response of other cases, becoming more sensitive to phosphorus. Tauc and optical sum-rule analyses further indicate that network topology defines the broader optical landscape, whereas doping primarily affects states near the gap and redistributes spectral weight. These findings establish defect topology as a microscopic control parameter that links amorphous structure, dopant activity, and optical response in amorphous silicon.
\end{abstract}

% \begin{keyword}
% dielectric function; Tauc analysis; Urbach-tail; optical gap; amorphous silicon; disorder; density functional theory
% \end{keyword}
% \date{\today}% It is always \today, today, %  but any date may be explicitly specified

\maketitle

% \section{\label{sec:Figures} Supporting Tables}

\begin{table}[!h]
\caption{
Structural characteristics of the undoped $a$-Si networks.
Coordination numbers were determined using a Si--Si cutoff of
$2.6$~\AA. Where $CN_n$ is the number of $n$-fold-coordinated
Si atoms, and $CN_{\mathrm{def}}$ is the total number of
non-fourfold-coordinated sites. $\langle\theta\rangle$ and
$\sigma_{\theta}$ are the mean Si--Si--Si bond angle and its
standard deviation, respectively.
}
\label{tab:structural_properties}
\centering
\footnotesize
\setlength{\tabcolsep}{12.5pt}
\begin{tabular}{lrrrrrrrrr}
\toprule
Model &
Density $(g/cm^3)$ &
$CN_3$ &
$CN_4$ &
$CN_5$ &
$CN_{\mathrm{def}}$ &
Defects (\%) &
$\langle\theta\rangle \pm \sigma_{\theta}$ ($^\circ$) \\
\midrule

R1 & 2.223 & 0 & 216 & 0  & 0  & 0.000  & $109.31 \pm 9.34$ \\

R2 & 2.232 & 6 & 210 & 0  & 6  & 2.778  & $109.24 \pm 9.66$ \\

R3 & 2.353 & 6 & 200 & 10 & 16 & 7.407  & $108.75 \pm 12.61$ \\

R4 & 2.316 & 4 & 206 & 6  & 10 & 4.630 & $109.14 \pm 7.60$ \\

\bottomrule
\end{tabular}
\end{table}

%%%%%%%%%%%%%%%%%%%%%%%%%%%%%%%   START   
%%%%%%%%%%%%%%%%%%%%%%%%%%%%%%%%TABLE 2%%%%%%%%%%%%%%%%%%%%%
\begin{table}[!t]
\caption{
Integrated dopant-induced dielectric perturbation for B- and P-doped $a$-Si models. 
$\mathcal{D}_{1}=\int|\Delta\varepsilon_{1}(E)|dE$ and 
$\mathcal{D}_{2}=\int|\Delta\varepsilon_{2}(E)|dE$ quantify changes in the real and imaginary dielectric responses, respectively, while
$\mathcal{D}=\int|\Delta\varepsilon(E)|dE$, with 
$|\Delta\varepsilon|=[(\Delta\varepsilon_{1})^{2}+(\Delta\varepsilon_{2})^{2}]^{1/2}$,
gives the total dielectric perturbation. Values are reported over $0$--$8$ eV and the visible region ($1.77$--$3.10$ eV).
}
\label{tab:dielectric_perturbation}
\centering
\footnotesize
\setlength{\tabcolsep}{12.5pt}
\begin{tabular}{llrrrrrr}
\toprule
& &
\multicolumn{3}{c}{$0$--$8$ eV} &
\multicolumn{3}{c}{Visible} \\
\cmidrule(lr){3-5}
\cmidrule(lr){6-8}

Model & Dopant &
$\mathcal{D}_{1}$ &
$\mathcal{D}_{2}$ &
$\mathcal{D}$ &
$\mathcal{D}_{1}$ &
$\mathcal{D}_{2}$ &
$\mathcal{D}$ \\
\midrule

\multirow{2}{*}{R1}
& B & 3.863 & 3.741 & 5.960 & 0.955 & 0.519 & 1.172 \\
& P & 3.802 & 3.459 & 5.651 & 0.852 & 0.656 & 1.134 \\

\midrule

\multirow{2}{*}{R2}
& B & 1.376 & 1.269 & 2.071 & 0.291 & 0.168 & 0.358 \\
& P & 0.816 & 0.830 & 1.322 & 0.255 & 0.211 & 0.370 \\

\midrule

\multirow{2}{*}{R3}
& B & 0.958 & 1.016 & 1.543 & 0.097 & 0.222 & 0.256 \\
& P & 1.742 & 1.861 & 2.894 & 0.439 & 0.308 & 0.589 \\

\midrule

\multirow{2}{*}{R4}
& B & 3.225 & 3.358 & 5.158 & 0.103 & 0.393 & 0.413 \\
& P & 2.440 & 2.318 & 3.734 & 0.305 & 0.259 & 0.445 \\
\bottomrule
\end{tabular}
\end{table}

% \clearpage
% \section{Supporting Figures}
\begin{figure}[!t]
    \centering
    \includegraphics[width=\linewidth]{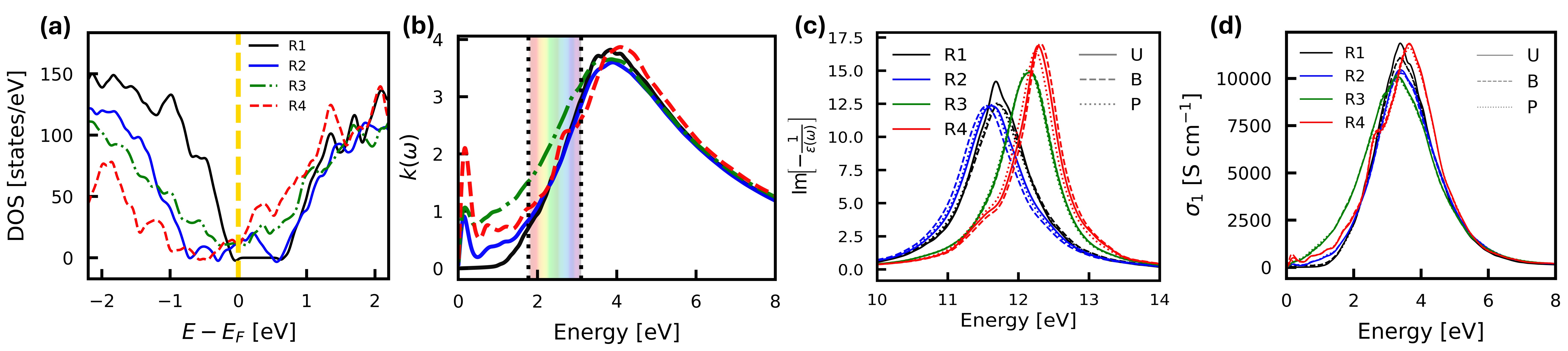}
    \caption{Electronic and optical responses of the four $a-$Si networks (R1-R4). (a) EDOS of undoped R1-R4, zeroed to the Fermi $E_{F}$. (b) Extinction coefficient $k(\omega)$ for the undoped, B- and P-doped. (c) EELF, $\mathrm{Im}[-1/\varepsilon(\omega)]$, (d) Real part of the optical conductivity $\sigma_1(\omega)$ . The lines (undoped (U), B-doped (B) and P-doped (P))}.
    \label{fig:Supplimentary_EDOS}
\end{figure}

\begin{figure*}{
    \centering
    \includegraphics[width=\linewidth]{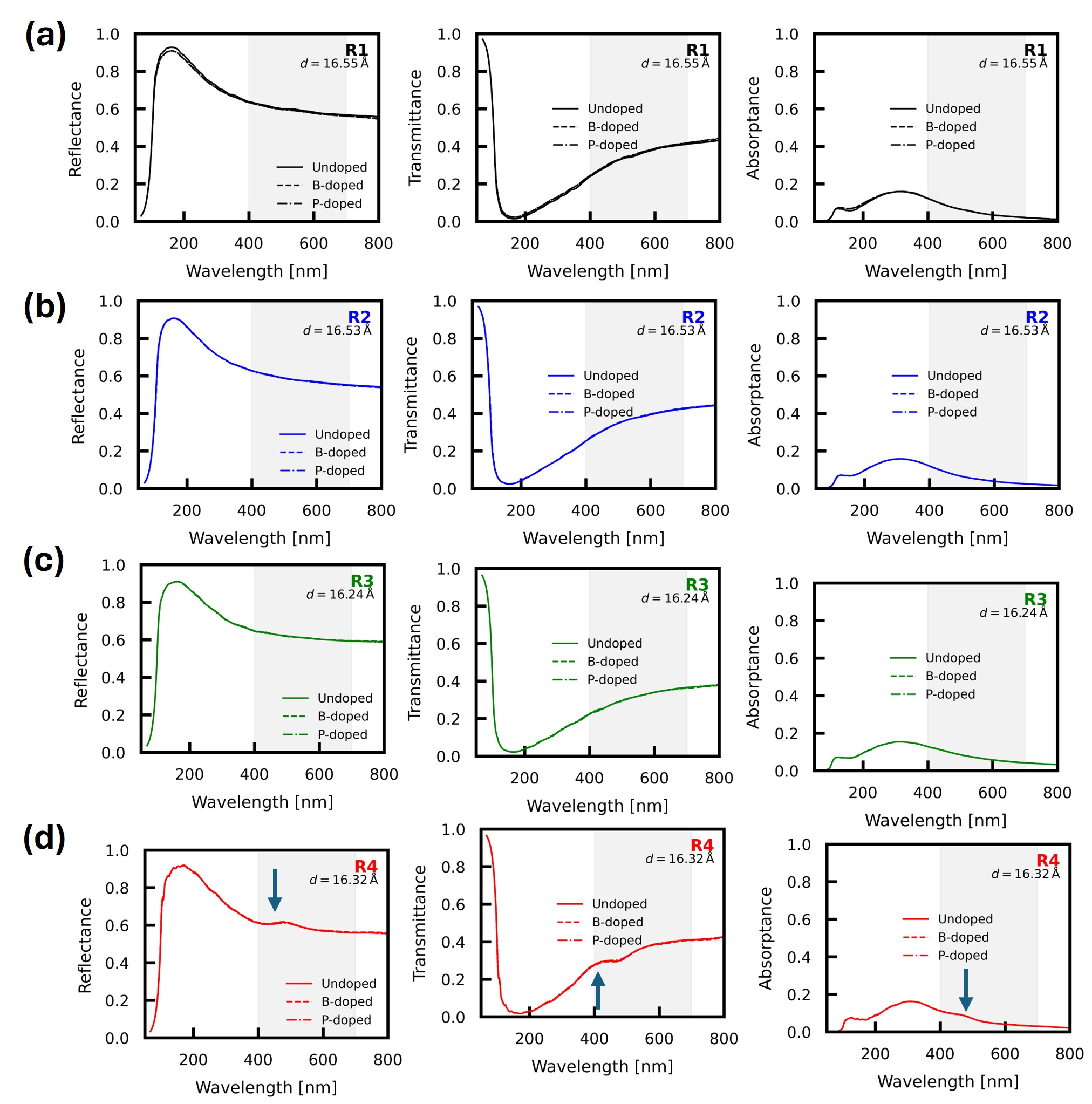}
    \caption{Reflectance, Transmittance and Absorptance of undoped, B-	 and P-doped $a-$Si against wavelength (a) R1, (b) R2 (c) R3 and (d) R4. The optical coefficients were determined using the respective simulation cell thickness. The gray shade depict the visible region $400-700$ nm. In R4, the arrow is used to indicate a change in the trend compared to the other models.}
}
    \label{fig:Supplimentary_ART}
\end{figure*}

\begin{figure}
{
    \centering
    \includegraphics[width=\linewidth]{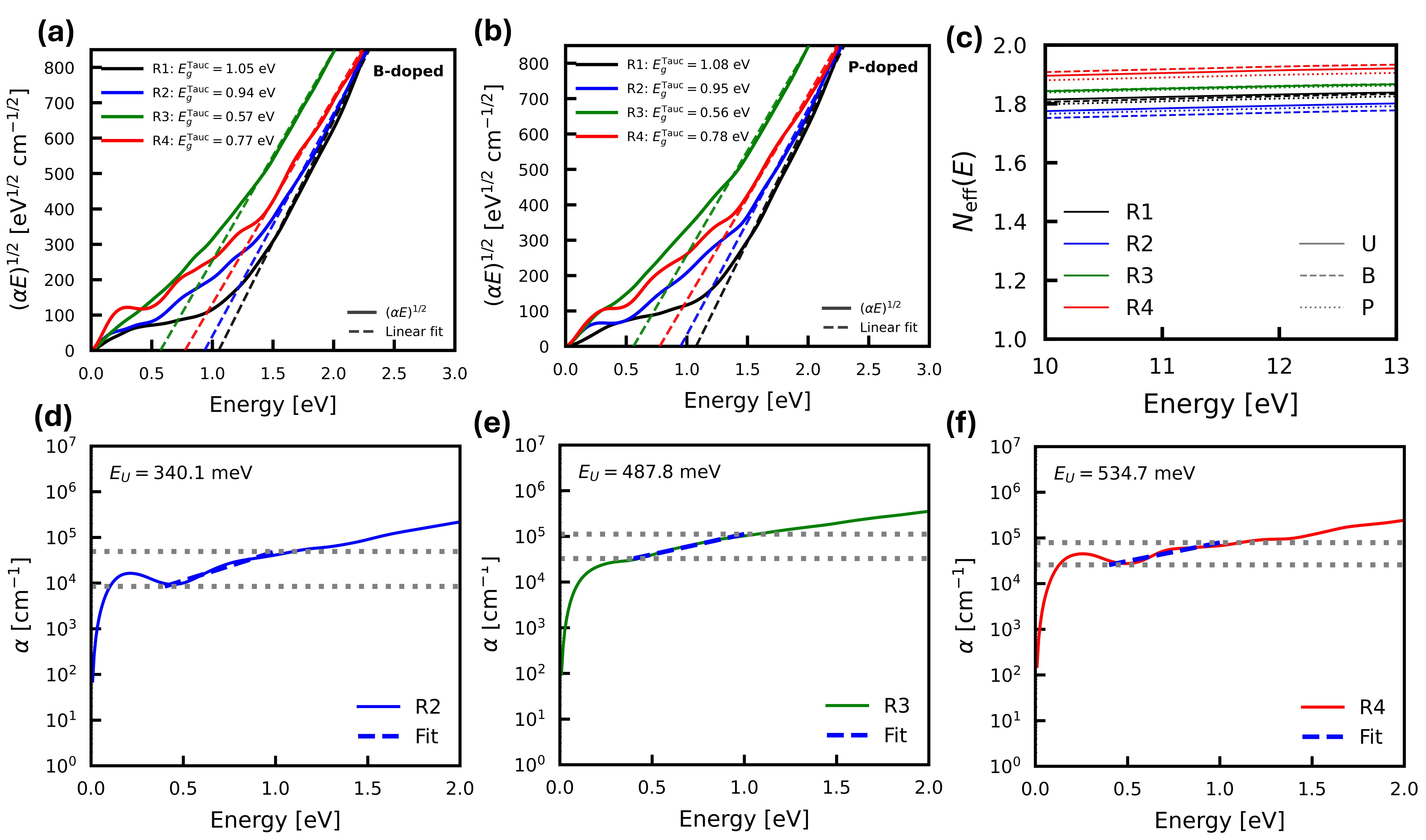}
    \caption{(a, b) Shown are the Tauc plots $(\alpha E)^{1/2}$ for the B-doped and P-doped configurations of R1-R4, with linear fits to extract the optical gaps. spectral weight in $10-13$ eV range for undoped ,B- and P-doped models. (d-f) Urbach analysis of R2 R3 and R4. Each panel shows the extracted Urbach energy $E_U$, within the Chosen fitting window.}
}
    \label{fig:}
\end{figure}

\bibliography{main_ref}% Produces the bibliography via BibTeX.